# 100-mJ class, sub-two-cycle, carrier-envelope phase-stable dual-chirped optical parametric amplification

Lu Xu,[1,2,†] Bing Xue,[1,2] Nobuhisa Ishii,[3] Jiro Itatani,[4] Katsumi Midorikawa,[2] Eiji J. Takahashi[1,5, *]

[1]Ultrafast Coherent Soft x-ray Photonics Research Team, RIKEN Center for Advanced Photonics, RIKEN, 2-1 Hirosawa, Wako, 3510198, Japan
[2]Attosecond Science Research Team, RIKEN Center for Advanced Photonics, RIKEN, 2-1 Hirosawa, Wako, 3510198, Japan
[3]Kansai Photon Science Institute, National Institutes for Quantum Science and Technology, Kizugawa, Kyoto 619-0215, Japan
[4]The Institute for Solid State Physics, The University of Tokyo, Kashiwanoha, 5-1-5, Kashiwa, Chiba 277-8581, Japan
[5]Extreme Laser Science Laboratory, RIKEN Cluster for Pioneering Research, RIKEN, 2-1 Hirosawa, Wako, 3510198, Japan
†lu.xu@riken.jp
*Corresponding author: ejtak@riken.jp

**Based on the dual-chirped optical parametric amplification and type-I BiB$_3$O$_6$(BiBO) crystals, the generation of >100 mJ, 10.4 fs, 10 Hz, carrier-to-envelope phase (CEP)-stable laser pulses, which are centered at 1.7 μm, is demonstrated; it produces a peak power of 10 TW. CEP-dependent high harmonic generation is implemented to confirm the sub-two-cycle pulse duration and CEP stabilization of infrared (IR) laser pulses. As far as we know, the obtained pulse energy and peak power represent the highest values for sub-two-cycle CEP-stable IR optical parametric amplification. Additionally, the prospects of achieving high-energy water window isolated attosecond pulses via our developed laser source are discussed.**

Currently, attosecond pulses, which were generated in an extreme ultraviolet (XUV)/soft X-ray photon energy region based on high harmonic generation (HHG) driven by femtosecond laser pulses had opened a new insight into electron dynamics and correlations in attosecond time scales [1]. To efficiently scale up both the output photon flux and photon energy of HHG had been one of the most exciting research topics in ultrafast laser science. According to the "cut-off" law of HHG explained with a semi-classical three-step model [2], it was evident that the driving laser with a longer wavelength was the essential requirement for extending the cut-off photon energy through ponderomotive scaling. Concurrently, the photon flux scaling law [3] indicated that the output photon flux rapidly decreased with the increasing driving wavelength [4].

With the use of a few-cycle laser field, the electron recombination in HHG was confined to a single wave-cycle, resulting in isolated attosecond pulse (IAP) emission [5]. In recent years, numerous studies had focused on generating soft X-ray harmonics, particularly in the water window region, that were driven by a few-cycle infrared (IR) laser source [6-10]. Additionally, previous experiments, where a few-cycle laser pulse with good stability in carrier-envelope phase (CEP) was generated via BiB$_3$O$_6$(BiBO) optical parametric chirped-pulse amplification (OPCPA) [11] and employed to demonstrate CEP-dependent HHG in the water window, revealed that a few-cycle pulse could trigger the quantum dynamics with sub-cycle accuracy and attosecond resolution owing to CEP stabilization of driving laser [6]. However, the output energy of IAP at the water window by IR laser had been restricted to below the pJ-class owing to the severe influence of HHG photon flux scaling [3] and the limitation of the driving energy of the IR pulses, indicating that the direct and efficient method of scaling up the photon flux of water window IAP must elevate the driving energy for HHG. Considering the foregoing, high energy, few-cycle, CEP-stable IR laser source was strongly desirable for achieving high photon flux and high photon energy IAP.

To develop a powerful IR source, a dual-chirped optical parametric amplification (DC-OPA) method was proposed [12] and experimentally demonstrated [13] as a variant of OPCPA. Previously, a 100-mJ, multi-cycle, CEP-unstable, DC-OPA laser [14] was developed and generated a nanojoule-level water window soft X-ray by HHG [15] based on the loosely focused geometry [16-19]. Our robust energy scaling method for HHG was expected to demonstrate single-shot absorption spectrum and live-cell imaging with a femtosecond time resolution. However, it was difficult to employ this DC-OPA laser for IAP energy scaling owing to the unstable CEP and multi-cycle pulse duration.

To the best of our knowledge, this letter was the first to report the simultaneous achievement of sub-two-cycle, 100-mJ class, CEP-stable IR laser pulses. By employing a joule-class Ti:sapphire pump laser, type-I $BiB_3O_6$ (BiBO) crystals and DC-OPA scheme, nearly one-octave bandwidth IR pulses with an energy of >100 mJ centered at 1.7 μm and a repetition rate of 10 Hz were generated. After compression in a bulk compressor, a pulse duration down to 10.4 fs was achieved to yield a peak power of 10 TW with shot-to-shot stability in CEP of 207 mrad rms. Furthermore, we performed a supercontinuum HHG experiment with Ar gas target under the neutral gas condition to demonstrate the few-cycle pulse duration and CEP stabilization of our developed DC-OPA source, where the half-cycle cutoff (HCO) was up to 210 eV based on the loosely focusing geometry. Finally, we discussed the prospects of a water window IAP source employing our novel laser source.

The system layout of this new DC-OPA was based on the multi-TW multi-cycle IR source reported elsewhere [20]. The main differences from the previously reported one were the following: (i) the 770 mJ Ti:sapphire pulses with a pulse duration of ~6 ps (positive-chirp) were applied as the pump for the DC-OPA. (ii) To further broaden the seed bandwidth, the pulses (4 mJ, 25 fs) from the frontend 1-kHz laser were injected into a 1.6 bar Kr gas cell for spectral broadening via optical filamentation [21]. Thereafter, difference frequency generation (DFG) with a 1-mm-thick BiBO crystal was employed to produce passively CEP-stable, broadband pulses (1.2 – 2.2 μm) with output energy of ~40 nJ. The crystal was cut at $\theta$ = 57.0 ° for type-II phase-matching in the x-z plane. Notably, the band stop mirror was employed to maximize the conversion efficiency in the DFG [22]. Using the acousto-optic programmable dispersive filter (AOPDF), the IR laser pulses with a pulse duration of ~4 ps (negative-chirp) were created as seed for the DC-OPA. Based on the calculated phase matching in type-I BiBO, only a part wavelength (740-802 nm) of our Ti:sapphire laser (740-840 nm) contributed to the amplification of the IR seed. Thus, the stretched pulse duration of the seed was shorter than the pump duration in the DC-OPA. (iii) Instead of a prism pair compressor used in the former [20], a bulk compressor was employed to avoid disturbing the CEP value that was passively stabilized during the seed generation. Moreover, a nitrogen-filled sealed box was used to cover the DC-OPA to avoid the absorption by water in the air.

Our DC-OPA comprised a three-stage configuration. It separated the total gain into three stages to avoid the loss of the conversion efficiency from the pump to the seed owing to the amplified parametric fluorescence, which inevitably appeared under the high pump intensity condition. The type-I BiBO crystals with a cutting angle ($\theta$) of 11 ° were employed in all the DC-OPA stages, and the thicknesses of the crystals at 1st, 2nd and 3rd stages were 6, 5, and 4 mm, respectively. By focusing the 1.5 mJ pump pulse with an intensity of 35 GW/cm² on the first DC-OPA, where the BiBO crystal was in the overlap of the Rayleigh length of the pump and the seed beams to maintain a good phase-matching condition, the injected seed pulses were amplified to 4 μJ. After collimating the beam to a diameter of 3 mm, the amplified seed pulses from the first DC-OPA were sent to the second DC-OPA, where they were subsequently amplified to 1.56 mJ with 30 mJ, 50 GW/cm² pump pulses that were collimated with a diameter of 3.5 mm. Afterward, the amplified seed pulses were collimated to match the BiBO crystal with a diameter of 15 mm at the third DC-OPA that was pumped by 740 mJ pulses with almost the same intensity as the one employed at the second DC-OPA. After the amplified seed pulses propagating ~1.2 m, they were spatially separated from the residual pump pulses. The final output energy of 105 mJ was achieved with a total conversion efficiency of ~13% from the pump to the seed. One of the crucial issues of the parametric amplification process was that the spatial profile of the amplified beam was extremely sensitive to that of the pump beam. Thus, the image of the pump beam was carefully relayed to each DC-OPA stage thereby maintaining smooth beam profiles (flat-top-like). Furthermore, the tilt angle of the collimated mirrors that were utilized in the IR laser system was elaboratively aligned to eliminate astigmatism in the IR laser pulses.

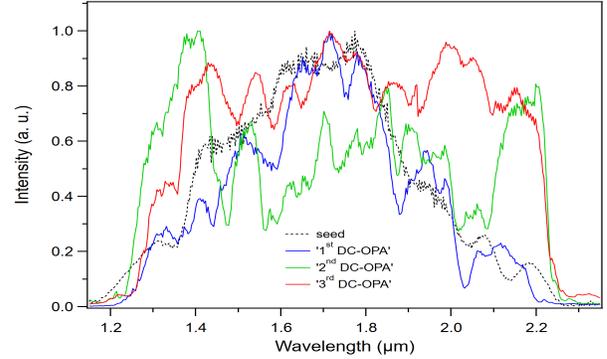

Fig. 1. Evolution of the seed spectrum: the injected seed spectrum (black dash-dotted line), the output spectrum of 1st DC-OPA stage (blue solid line), the output spectrum of 2nd DC-OPA stage (green solid line), and the output spectrum of 3rd DC-OPA stage (red solid line).

Concurrently, by adjusting the noncollinear angle (~0.6°) between the pump and seed introduced at each DC-OPA stage, the first DC-OPA stage concentrated on the amplification of the center of the injected spectrum (blue solid line in Fig. 1), and the second DC-OPA stage focused the amplification of the edge part of the output spectrum from the first DC-OPA stage (green solid line in Fig. 1). These optimizations ensured that an approximately octave output bandwidth was obtained after the final DC-OPA stage (red solid line in Fig. 1). Moreover, we added a hole in the injected seed spectrum by AOPDF to avoid the saturation around the center wavelength, because it induced a complex compressed spectral phase distortion via the reversed energy flow between the pump and seed pulses in the optical parametric process [20,23].

After the amplification, the IR laser pulses were expanded to a diameter of 60 mm and temporally compressed with a 130-mm-long water-free fused silica bulk compressor to obtain an output energy of 102 mJ (the throughput of the compressor was 97%). To characterize the duration of the compressed pulse, second harmonic generation frequency-resolved optical gating (SHG-FROG) was adopted for the IR laser pulses. The spectral phases of the IR laser pulses were retrieved from SHG-FROG traces and fed back to AOPDF. After performing several iterations to optimize the spectral phase, the fully compressed IR laser pulses with a nearly flat spectral phase were obtained. The solid red line and dashed black line in Fig. 2(e) showed the spectrum and the spectral phase, which were retrieved from the measured SHG-FROG trace expressed in Fig. 2(a), respectively. The corresponding temporal pulse profile of the retrieved pulses with a pulse duration of 10.4 fs (FWHM) was shown in Fig. 2(d), which corresponded to a 1.8-cycle duration of the electric field at a center wavelength of 1.7 μm, with the reconstruction error of 0.9%. Notably, because the total optical path of the seed pulses in this high output energy DC-OPA system was ~16 m, absorption by water in the experimental environment caused significant spectral phase distortion in the IR laser pulses. As demonstrated in Fig. 2(c), the distortion around 1.4 and 1.9 μm disturbed the measured SHG-FROG trace, thus causing the reconstructed spectral phase to lose sufficient accuracy to be fed back to AOPDF for complete compression. Therefore, as aforementioned, a sealed box with a nitrogen density of >99 % and humidity of <5 % was installed to cover the DC-OPA module to maintain the sub-two cycles output of the IR laser system.

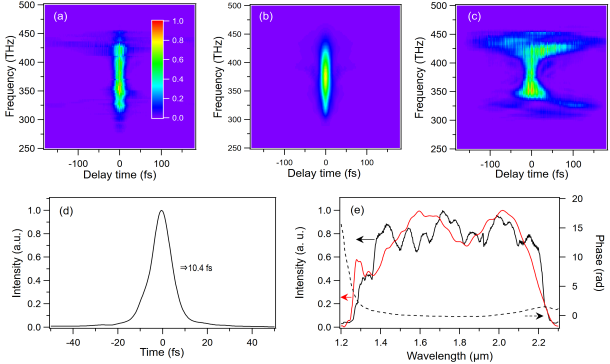

Fig. 2. SHG-FROG results of the DC-OPA output: (a) measured SHG-FROG trace, (b) retrieved SHG-FROG trace with an error of 0.9 %, (c) measured SHG-FROG trace without the sealed $N_2$ box, (d) reconstructed temporal profile with a pulse duration of 10.4 fs (FWHM), (e) reconstructed spectrum (red solid line), reconstructed spectral phase (black dashed line) and measured spectrum after the third DC-OPA stage (black solid line).

To characterize the CEP stability of the IR laser system, a home-built single-shot f-to-2f interferometer was implemented after the bulk compressor. A part of the compressed IR pulses was used to generate a white-light continuum in a $CaF_2$ plate followed by a BBO crystal, where the longer wavelength components in the continuum were frequency-doubled and interfered with the shorter wavelength components through a polarizer to generate the interference fringe ranging from 820 to 880 nm. To qualitatively analyzed and discussed the influence of the IR energy fluctuation on the f-to-2f interferometer, the leak energy of the output IR laser pulses was simultaneously recorded to check the correlation between the CEP shift and energy fluctuation [24].

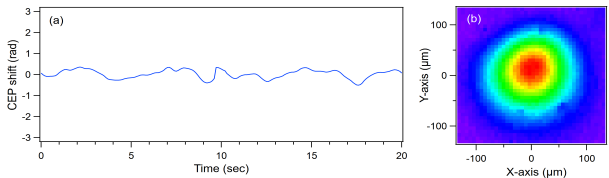

Fig. 3. Single-shot CEP measurements in the f-to-2f interferometer: (a) CEP shift in 20 s, and (b) focusing beam profile measured with a CCD camera.

As shown in Fig. 3(a), the variation of the CEP shift calculated from collected interferograms was 207 mrad rms, and the correlation coefficient of the energy fluctuation and CEP shift was 0.15. Further, we verified focusability using a 2-m focusing geometry. Figure 3(b) showed the measured beam profile (~200 μm, $1/e^2$) on the focus plane with a beam diameter of 60 mm. By recording the beam size of the focusing beam profile up to four times the Rayleigh length before and after the focus plane, the obtained $M^2$ values of the focused beam profile in the horizontal and vertical directions were 1.22 and 1. 31, respectively. The beam of such quality was applicable to most of the strong-field experiments such as HHG.

To verify the sub-two-cycle duration and CEP stabilization of the output IR laser pulses, high-order harmonics (HHs) were generated and analyzed. The IR laser beam whose pulse energy was adjusted by an aperture was focused on the Ar gas cell with a 2-m focusing geometry. The gas cell had pinholes at both ends of a 30-mm-long squeezed tube for the focused IR beams to pass through. The high harmonic spectrum was measured by a grating-based spectrometer (1200 lines/mm) with a microchannel plate with a fluorescent screen that was imaged and recorded by a CCD camera, which was ~1.5 m away from the gas cell. The details of the HHG setup and measurement system had been reported elsewhere [15].

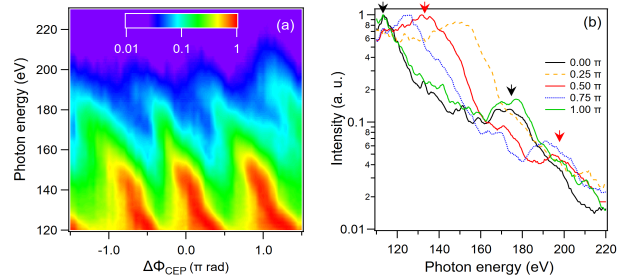

Fig. 4. Measurements of HHG spectrum generated from Ar with the backing pressure of 900 Torr and injected IR laser pulse energy of 5 mJ. (a) Experimentally measured HHG spectrum recorded at the relative CEP value ($\Delta\Phi_{CEP}$) in steps of $0.05\pi$ rad; (b) CEP-dependent HCO associated with the electron trajectories in a sub-two-cycle pulse, where the HHG spectrum is recorded at $\Delta\Phi_{CEP}$ in steps of $0.25\pi$ rad.

The generated HHs were measured at different backing pressures from 700 to 950 torr in steps of 25 torr, by adjusting the injected energy of the IR laser pulses with the aperture. The harmonic yield gradually increased at <900 torr with the increasing gas pressure and decreased when the backing pressure exceeded 900 torr. The harmonic yield exhibited the quadratic dependence of the harmonic yield as a function of the gas backing pressure below 900 torr, indicating that the conversion efficiency and the photon flux of HHG depended on the Ar gas pressure. As a result, the best phase matching of HHG with 2-m focusing was achieved with the injected energy of 5 mJ adjusted by an 18 mm beam diameter, and the effective gas pressure at the interaction region of 90 Torr estimated from the backing pressure [25]. Additionally, the beam divergence of HHs was improved at the optimized phase matching condition. Meanwhile, the geometrical focus intensity in a vacuum was estimated to reach $10^{15}$ (W/cm$^2$) under the experimental conditions. This estimated focus intensity would break the phase-matching condition for a neutral atom. Conversely, we estimated the effective interaction intensity of the pump pulse from the cutoff law [2]. From the observed Ar HHG spectral distribution in Fig. 4 (b), the interaction intensity was estimated to be $2.3 \times 10^{14}$ (W/cm$^2$) which gave the estimated ionization probability of less than 1% from the Ammosov-Delone-Krainov model [26]. This low ionization was supported by the quadratic increase in the harmonic intensity with the Ar gas pressure, in which phase matching was satisfied in the nonionized condition. The almost perfect Gaussian profile of the HH also indicated that there was no density disturbance due to ionization in the interaction region. Notably, the laser pulse was focused by the plano-convex lens in this HHG experiment. The focus plane position of each driving wavelength was not exactly the same along the propagation direction because our laser pulse exhibited a one-octave spectrum. Therefore, the injected whole IR laser could not be exactly focused on the single focusing point, and this decreased the effective focusing intensity. It will be improved with a concave or parabolic mirror focus scheme in the future.

After the optimization of the phase matching, the CEP dependence of HHG was demonstrated by recording HHG spectra (the exposure time of the CCD was set to be 1 sec.) at relative CEP values in steps of $0.05\pi$ rad. In Fig. 4(a), similar intensity distributions of the HHG spectrum repeated every $\pi$ rad of CEP shift since the spatial inversion of the electric field of the driving pulses did not affect the spectral intensity of HHG. It exhibited an explicit dependence on the CEP shift of the driving IR laser pulses. Furthermore, the HHG spectra were recorded at five relative CEP values in steps of $0.25\pi$ rad to characterize the sub-two-cycle pulse duration of the IR laser. As demonstrated in Fig. 4(b), one or two peaks appeared in the HHG spectrum depending on the CEP shift, which was the consequence of the HCO that originated from several electron trajectories in the HHG process driven by a few-cycle pulse [27]. To elucidate the evolution of the recorded CEP-dependent HCO spectrum in Fig. 4(b), we referred to the *cos-like* and *sin-like* waveform models of a sub-two-cycle pulse in Ref. [6]. According to the experimental results in Fig. 4(b), the HHG spectrum with two peaks (marked by black or red arrows) driven by a near *sin*-like waveform (black or green solid line) evolved to the HHG spectrum that was driven by a near *cos*-like waveform (red solid line) when the CEP was shifted by $0.5\pi$ rad with the cut-off energy shifted to a higher photon energy (up to 210 eV). This drastic cut-off energy shift of the HCO indicated that the HHG process was driven by a sub-two-cycle laser pulse [6].

Finally, taking our developed 100-mJ class a few-cycle laser results into account, we can design a high-energy water window IAP source and discuss its expected pulse energy. Based on the energy scaling in the phase-matched condition [28], a 7.6-m-long loosely focusing geometry with a beam size of 60 mm was employed in our design to take advantage of the whole output energy of the IR laser system [15], where the focused beam diameter was 600 μm. The loosely focused strategy [28] to scale up the continuum harmonic energy can well work under a few-cycle driving laser [29,30]. Concurrently, the interaction gas pressures were 40 torr and 350 torr to compensate for the Gouy phase shifts for Ne and He, respectively. By employing the 1.7 μm, sub-two-cycle driving laser with peak intensities of $4\times10^{14}$ W/cm$^2$ and $5.6\times10^{14}$ W/cm$^2$ for Ne and He, respectively, the corresponding cut-off photon energies were expected up to ~370 eV for Ne and ~510 eV for He, where a stable phase-matching was maintained in the neutral gas condition because of the relatively low ionization degree. As the requirement of the absorption-limited condition [31], the gas medium length should be 60 mm and 150 mm for Ne and He, respectively, which were within the Rayleigh length (170 mm) of our designed geometry. Based on the photon flux scaling law [3] and the previously reported conversion efficiency [15], the expected harmonic energies from the carbon K-shell edge (284 eV) to the cutoff region (water window) will be ~15 nJ in Ne and ~5 nJ in He respectively. Thanks to the nano-joule class IAP, its peak power will be a sub-GW class with a duration of below 100 as [9].

In conclusion, we demonstrated a 100 mJ-class, 10.4 fs, 10 Hz, type-I BiBO-based CEP-stable (207 mrad rms) DC-OPA IR laser source, resulting in peak power of 10 TW. The establishment of CEP stabilization and the sub-two-cycle pulse duration of the IR laser source was revealed from the evolution of the CEP-dependent HHG and the CEP-dependent HCO process (up to 210 eV) in the observed HHG spectra. Furthermore, the prospect of a high energy IAP in the water window soft X-ray that is driven by our developed laser source was discussed. Thanks to the nano-joule class soft x-ray HHG by our laser source, a peak power of IAP will reach sub-GW class, and both the single-shot and attosecond temporal resolution will be provided to the applications of ultrafast soft x-ray science.

**Funding.**
Ministry of Education, Culture, Sports, Science and Technology of Japan (MEXT) through Grants-in-Aid under Grant Nos. 21H01850 and 19H05628, in part, by the MEXT Quantum Leap Flagship Program (Q-LEAP) (Grant No. JP-MXS0118068681).

**Acknowledgments.**
We thank Dr. Y. Fu and Dr. K. Nishimura for useful comments on the HHG experiment.

**Disclosures.** The authors declare no conflicts of interest.

**Data availability.** Data underlying the results presented in this paper are not publicly available at this time but may be obtained from the authors upon reasonable request.